\documentclass[letterpaper,titlepage,11pt]{article}
\usepackage{hyperref}

\usepackage{amssymb,amsmath,amsfonts}
\usepackage{epsfig}

\def\clock{{\count0=\time
           \divide\count0 60
           \ifnum\count0<10 0\fi\the\count0
           \multiply\count0 -60 \advance\count0 \time
           :\ifnum\count0<10 0\fi \the\count0
         }}
\newcommand{\timestamp}{{\small\vbox{\hbox{\tt\jobname.tex}
\hbox{\the\day/\the\month/\the\year, \clock}}}}

\newcommand{\be}{\begin{equation}} \newcommand{\ee}{\end{equation}}
\newcommand{\bea}{\begin{eqnarray}} \newcommand{\eea}{\end{eqnarray}}

\newcommand{\CT}{\mathcal{T}}

\newcommand{\CM}{\mathcal{M}}

\newcommand{\id}{\hbox{1\kern-.27em l}}
\newcommand{\sid}{\hbox{\scriptsize1\kern-.27em l}}

\newcommand{\we}{\kern-.1em\wedge\kern-.1em}
\newcommand{\scal}{\kern-.13em\cdot\kern-.13em}

\newcommand{\II}{I\kern-.09em I}

\newcommand{\R}{\mathbb{R}}
\newcommand{\T}{\mathbb{T}}

\newcommand{\nn}{\nonumber}

\newcommand{\spa}{\ , \ \ }

\numberwithin{equation}{section}
\begin{document}

\begin{titlepage}

\rightline{\vbox{\small\hbox{\tt hep-th/0403103} }}
\vskip 3cm

\centerline{\Large \bf General Definition of Gravitational Tension}
%\vskip 0.1cm
%\centerline{\Large \bf }

\vskip 1.6cm
\centerline{\bf Troels Harmark and Niels A. Obers}
\vskip 0.5cm
\centerline{\sl The Niels Bohr Institute}
\centerline{\sl Blegdamsvej 17, 2100 Copenhagen \O, Denmark}

\vskip 0.5cm

\centerline{\small\tt harmark@nbi.dk, obers@nbi.dk}

\vskip 1.6cm

\centerline{\bf Abstract} \vskip 0.2cm \noindent
In this note we give a general definition of
the gravitational tension in a given asymptotically 
translationally-invariant spatial
direction of a space-time.
The tension is defined via the
extrinsic curvature in analogy with the Hawking-Horowitz
definition of energy. We show the consistency with the ADM tension
formulas for asymptotically-flat space-times, in particular for
Kaluza-Klein black hole solutions.
Moreover, we apply the general tension formula to near-extremal branes,
constituting a check for non-asymptotically flat space-times.

%\vskip 0.5cm
%\leftline{\timestamp}

\end{titlepage}

%\pagestyle{empty}
%\tableofcontents
%\newpage

\pagestyle{plain}
\setcounter{page}{1}

%%%%%%%%%%%%%%%%%%%%%%%%%%%%%%%%%%%%%%%%%%%%%%%%%%%%%%%%%%%%%%
\section{Introduction and summary}

In this note we give a general definition of the gravitational tension
in a given asymptotically translationally-invariant
spatial direction of a space-time. The definition is
general in the sense that while it reduces to the correct
tension formulas
for asymptotically-flat space-times it can also be applied to
non-asymptotically flat space-times.
So far, general tension formulas are known for asymptotically-flat
space-times
\cite{Traschen:2001pb,Townsend:2001rg,Harmark:2003dg,Kol:2003if} but
it has only been computed in particular cases of
non-asymptotically flat space-times
\cite{Myers:1999ps,Awad:1999xx,DeBenedictis:1999yn,Cai:1999xg}.

The general definition of gravitational tension is
obtained in Section \ref{sectens}
by generalizing the derivation of the Hawking-Horowitz
energy formula \cite{Hawking:1996fd}.
In particular this means that
the tension is defined relative to a reference space-time,
thus making it a necessary requirement
for the applicability of the tension definition
to have a sensible choice of
reference space-time.
The tension formula is derived by applying the Hamiltonian
formalism to a foliation of the space-time
along the spatial direction in which we want to measure the
tension, in analogy with the Hawking-Horowitz energy formula
in which the space-time is foliated along the time-direction.%
\footnote{Note that the idea of defining the tension using the
Hamiltonian formalism for a spatial direction is from
\cite{Traschen:2001pb} where it is used to derive a formula for
the tension for asymptotically-flat space-times.}

In Section \ref{secflat} we show that the general tension
formula correctly reduces to the known ADM-type expressions
\cite{Traschen:2001pb,Townsend:2001rg,Harmark:2003dg,Kol:2003if}
for asymptotically-flat space-times. Moreover, following the work of
\cite{Traschen:2001pb,Townsend:2001rg,Harmark:2003dg,Kol:2003if}, we present
a general Smarr formula that holds in this case and gives
the first law of thermodynamics, including the work term associated
to the tension. We also give the general bounds satisfied by
the tension following from the positivity of gravitational tension
\cite{Traschen:2003jm,Shiromizu:2003gc} and the Strong Energy Condition.

As an illustration of the general tension formula we have included
two simple examples in Section \ref{secex}. The first example
is that of vacuum solutions of higher-dimensional
General Relativity for Kaluza-Klein
type space-times, i.e. Minkowski-space times a circle.
Of specific interest are static and neutral solutions with event horizons,
which we here call Kaluza-Klein black holes.
We show that one recovers the tension formula for the
circle direction in this case, previously
obtained in \cite{Harmark:2003dg,Kol:2003if}
and used for Kaluza-Klein black holes
in 
%\cite{Harmark:2003dg,Kol:2003if,Harmark:2002tr,Harmark:2003eg,Sorkin:2003ka,Kudoh:2003ki,Harmark:2003yz,Sorkin:2004qq}.
\cite{Harmark:2003dg,Kol:2003if,blob1}.

The second example we consider is that of near-extremal branes
of type IIA/B String theory and M-theory, which are non-asymptotically
flat space-times. We show that
the general expression correctly reproduces the known results
\cite{Myers:1999ps,Cai:1999xg}
for the tension along the spatial world-volume directions.
For most of these cases the measured tension is negative 
which is natural since
the gravitational tension has a dual interpretation as the
pressure of the non-gravitational theory living on the world-volume
of the brane.

Since branes and other objects with tensions
play a central role in General Relativity, String/M-theory
and holographic dualities between gravitational and non-gravitational
theories,
many other applications of the general gravitational tension formula
can be envisioned.
In a forthcoming work \cite{Harmark:2004}, we consider a combination of
the two examples mentioned above, namely near-extremal branes on a
transverse circle.
This will constitute a new application of the definition of gravitational tension
given in this note.

Finally, we have included three appendices. 
In the first one we show how to get the ADM-type
tension formula for asymptotically flat space-times from
the general tension formula.
In the
second one we solve the linearized Einstein equations to show
the connection between the ADM-type expressions
for asymptotically-flat space-times and a second set
of expressions that can be derived in this case.
The third appendix contains some details
of the proof of the Smarr formula.

%%%%%%%%%%%%%%%%%%%%%%%%%%%%%%%%%%%%%%%%%%%%%%%%%%%%%%%%%%%%%%
\section{General definition of gravitational tension \label{sectens} }

In Section \ref{hamapp} we use the Hamiltonian approach
for a foliation of the space-time along a spatial direction
to argue for the general definition of gravitational tension
for a space-time. In the process we repeat the derivation
of the Hawking-Horowitz formula for energy \cite{Hawking:1996fd}.
In Section \ref{genform} we summarize the results of Section
\ref{hamapp} and write down the explicit expression for the
definition of gravitational tension.

%%%%%%%%%%%%%%%%%%%%%%%%%%%%%%%%%%%%%%%%%%%%%%
\subsection{Hamiltonian approach to definition of energy
and tension}
\label{hamapp}

In this section we generalize the Hamiltonian approach
to defining energy in
\cite{Hawking:1996fd} so that we also get a general definition of the
tension.
In order to be pedagogical and complete part of the following derivation is
a repeat of \cite{Hawking:1996fd}.
See also \cite{Wald:1984rg} for the formalism used.

The basic idea in the following is to generalize the Hamiltonian
approach of \cite{Hawking:1996fd} so that we can consider both
a foliation of the space-time along a time direction and a
spatial direction. While for a time direction we get the Hawking-Horowitz
definition of energy we get for a spatial direction instead a
general definition of tension. The idea of defining the tension
from the Hamiltonian along a spatial direction is
from \cite{Traschen:2001pb} where it is used
to get a general definition of the tension
for asymptotically-flat space-times.

The Einstein-Hilbert action for a $D$-dimensional Lorentzian manifold
$(M,g_{\mu \nu})$ with boundary $\partial M$ is
\begin{equation}
\label{Mact}
I = \frac{1}{16\pi G} \int_M R + \frac{1}{8\pi G} \int_{\partial M} K \ ,
\end{equation}
where $g_{\mu \nu}$ is the metric on $M$.
Here $R$ is the Ricci scalar curvature and $K$ is the extrinsic curvature
on $\partial M$. We suppress the integration measures here and
in the following.
We could also add matter terms to the action. However, since they in general
only include first order derivatives, these will not
be important for defining the energy or tension and we thus ignore them
in this discussion.

In the following we shall use \eqref{Mact} to define a Hamiltonian
on $M$ with respect to a given time-like or space-like foliation
of $M$. Subtracting the Hamiltonian on a reference space then gives a
natural and general definition of the value of the Hamiltonian
which we subsequently use to define the energy and tension.

Consider a family of codimension one submanifolds $\{ \Sigma_x \}$ of $M$
labeled by
the parameter $x \in [x_{\rm min} , x_{\rm max}]$,
given such that $\{ \Sigma_x \}$ is a foliation of $M$.
Consider moreover a vector field $X^\mu$ on $M$ satisfying
$X^{\mu} \partial_\mu x = 1$.
We can think of $X^\mu$ as defining the ``flow of $x$'' throughout the
space-time. In the following we consider either the case where
the family of submanifolds $\{ \Sigma_x \}$ all are Euclidean and $X^\mu$
is time-like (in which case
$x$ can be thought of as a time-coordinate) or the case where
$\{ \Sigma_x \}$ all are Lorentzian and $X^\mu$ is space-like.

Consider now the unit normal $n^\mu$ of the submanifold $\Sigma_x$,
so that $n^\mu n_\mu = s$ where
$s=1$ ($s=-1$) if $X^\mu$ is space-like (time-like).
We can then decompose $X^{\mu}$ into parts
normal and tangential to $\Sigma_x$
\begin{equation}
X^{\mu} = N n^{\mu} + N^{\mu}  \ ,
\end{equation}
defining the lapse function $N$ and shift vector $N^{\mu}$.
The metric on $\Sigma_x$ with respect to $n^\mu$ is
\begin{equation}
\label{defgx}
g^{(x)}_{\mu \nu} = g_{\mu \nu} - s \, n_{\mu} n_{\nu} \ .
\end{equation}
The extrinsic curvature tensor on $\Sigma_x$ with respect to $n^\mu$ is
\begin{equation}
K_{\mu \nu} = D_\mu n_\nu \ .
\end{equation}

We assume the foliation $\{ \Sigma_x \}$ of $M$ to be such that
the boundary $\partial M$ of $M$ consist of three separate pieces:
The ``initial'' boundary $\Sigma_{x_{\rm min}}$, the ``final'' boundary
$\Sigma_{x_{\rm max}}$,
and the ``asymptotic boundary'' $\Sigma^\infty$, so that
\begin{equation}
\label{partM}
\partial M = \Sigma_{x_{\rm min}} \cup \Sigma_{x_{\rm max}}
\cup \Sigma^\infty \ .
\end{equation}
Let $r^\mu$ be a unit normal vector field on $\Sigma^\infty$.
We assume furthermore the  foliation $\{ \Sigma_x \}$ of $M$ to be such that
$n^\mu$ is a tangent on $\Sigma^\infty$, i.e. such that $r \cdot n = 0$.
Defining now the intersections $S^\infty_x = \Sigma_x \cap \Sigma^\infty$
we can foliate $\Sigma^\infty$ by the family of submanifolds
$\{ S^\infty_x \}$.

Note that the data of the metric $g_{\mu \nu}$ is contained
in $(g^{(x)}_{\mu \nu},N, N_\mu)$ and vice versa. To obtain the Hamiltonian
corresponding to the flow of $x$, we now rewrite the action \eqref{Mact}
in terms of the latter.

Defining $G_{\mu \nu} = R_{\mu \nu} - \frac{1}{2} g_{\mu \nu} R$
we have
\begin{equation}
R = 2 s (R_{\mu \nu} - G_{\mu \nu} ) n^\mu n^\nu
\spa
2 G_{\mu \nu} n^\mu n^\nu = - s R^{(x)} + K^2 - K_{\mu \nu} K^{\mu \nu} \ ,
\end{equation}
\begin{equation}
R_{\mu \nu} n^\mu n^\nu =K^2 - K_{\mu \nu} K^{\mu \nu}
 - D_\mu ( n^\mu D_\nu n^\nu )
+  D_\nu ( n^\mu D_\mu n^\nu ) \ ,
\end{equation}
so that
\begin{equation}
\label{Rexp}
R =R^{(x)} + s [ K^2 - K_{\mu \nu} K^{\mu \nu} ]
-2 s [ D_\mu ( n^\mu D_\nu n^\nu )
- D_\nu ( n^\mu D_\mu n^\nu ) ] \ ,
\end{equation}
for a given submanifold $\Sigma_x$.
The last two terms here are boundary terms and combining with
the extrinsic curvature term we get the boundary action
\begin{equation}
\label{bound1}
8 \pi G \, I_{\rm bd} =
\int_{\partial M} K  - s \int_{M} [  D_\mu ( n^\mu D_\nu n^\nu)
- D_\nu ( n^\mu D_\mu n^\nu) ] \ .
\end{equation}
For the contribution on the $\Sigma_{x_{\rm min}}$ and
$\Sigma_{x_{\rm max}}$ part of $\partial M$ (with normal
vectors $- n^\mu$ at $\Sigma_{x_{\rm min}}$ and
$n^\mu$ at $\Sigma_{x_{\rm max}}$) we have
\begin{equation}
- \int_{\Sigma_{x_{\rm min}}} [ g^{\mu \nu} D_\mu n_\nu - s (n\cdot n)
D_\nu n^\nu ]
+ \int_{\Sigma_{x_{\rm max}}} [ g^{\mu \nu} D_\mu n_\nu - s (n\cdot n)
D_\nu n^\nu ]= 0 \ .
\end{equation}
Here the last term in \eqref{bound1} does not contribute since
$n_\nu n^\mu D_\mu n^\nu = 0$ because $n$ is unit vector.

For the contribution from the boundary at infinity $\Sigma^\infty$
(with normal vector $r^\mu$) we have on the other hand
\begin{equation}
\int_{\Sigma^\infty} [g^{\mu \nu} D_\mu r_\nu + s \,
r_\nu n^\mu D_\mu n^\nu ] \ ,
\end{equation}
since the second term in \eqref{bound1} does not contribute because
$r\cdot n =0$.
After integrating by parts the second term in this expression
we can write this as
\begin{equation}
\label{bt1}
\int_{\Sigma^\infty} [g^{\mu \nu} - s \, n^\mu n^\nu] D_\mu r_\nu \ .
\end{equation}
We recognize now $g^{(x)}_{\mu \nu} =g_{\mu \nu} - s \, n_\mu n_\nu$ as the
metric on $\Sigma_x$ 
so that the integrand of \eqref{bt1}
 is nothing but the
extrinsic curvature $K^{(D-2)} = (g^{(x)})^{\mu \nu} D^{(x)}_\mu r_\nu$ 
of the surface $S^\infty_x$  in $\Sigma_x$.
Here $D^{(x)}_\mu$ is the covariant derivative in $\Sigma_x$.
We get thus that \eqref{bound1} can be written as
\begin{equation}
\label{bound2}
8 \pi G \, I_{\rm bd} = \int dx N \int_{S_x^{\infty}}  K^{(D-2)} \ .
\end{equation}
Using the above we can therefore write the action \eqref{Mact} as
\begin{equation}
\label{xact}
I = \frac{1}{16 \pi G} \int dx N \left[
\int_{\Sigma_x} (R^{(x)} + s [K^2 - K_{\mu \nu} K^{\mu \nu} ])
+ 2 \int_{S^\infty_x} K^{(D-2)} \right] \ .
\end{equation}

We now further rewrite the bulk part of \eqref{xact}
\begin{equation}
\mathcal{L}_{\rm bulk}
= \sqrt{ g^{(x)}} \left(R^{(x)} + s [ K^2 - K_{\mu \nu} K^{\mu \nu} ]
\right) \ ,
\end{equation}
to a form appropriate for extracting the Hamiltonian.
For this we introduce the canonical momentum
\begin{equation}
p^{\mu \nu} \equiv \frac{1}{\sqrt{g^{(x)}}} \pi^{\mu \nu}
= \frac{1}{\sqrt{g^{(x)}}}
\frac{\partial {\cal{L}}_{\rm bulk}}{\partial \dot{g}^{(x)}_{\mu \nu}} \ ,
\end{equation}
with
\begin{equation}
\dot{g}^{(x)}_{\mu \nu} = (g^{(x)})_\mu^{\ \rho} (g^{(x)})_\nu^{\ \sigma}
\mathcal{L}_X g_{\rho \sigma} \ ,
\end{equation}
and use that the extrinsic curvature $K_{\mu \nu}$ is related
to $\dot{g}^{(x)}_{\mu \nu}$ by
\begin{equation}
K_{\mu \nu} = \frac{1}{2N} \left[ \dot{g}^{(x)}_{\mu \nu}
- 2 D^{(x)}_{(\mu}N_{\nu)} \right] \ .
\end{equation}
Then we can write
\begin{equation}
\label{bulk2}
\mathcal{L}_{\rm bulk}
= \sqrt{ g^{(x)}} \left( p^{\mu \nu} \dot{g}^{(x)}_{\mu \nu}
- [-s N {\cal{H}} + N^{\mu} {\cal{H}}_\mu ]
- 2 D_\mu (N_\nu p^{\mu \nu})\right) \ ,
\end{equation}
where the Hamiltonian constraints are given by
\begin{equation}
{\cal{H}} =R^{(x)} + p^{\mu \nu} p_{\mu \nu} - \frac{1}{D-2} p^2
\spa
{\cal{H}}_\mu = - 2 D_{\mu} p^{\mu \nu} \ ,
\end{equation}
with $p = p_\mu{}^\mu$. The last term in \eqref{bulk2} will contribute
an extra boundary term giving the action
\begin{equation}
I = \frac{1}{16 \pi G} \int dx \left[
\int_{\Sigma_x} \left( p^{\mu \nu} \dot{g}^{(x)}_{\mu \nu}
- [-s N {\cal{H}} + N^{\mu} {\cal{H}}_\mu ] \right)
+ 2 \int_{S^\infty_x} \left( N K^{(D-2)} - N^\nu p_{\mu \nu} r_{\nu} \right)
\right] \ .
\end{equation}
{}From this action we can extract the Hamiltonian
\begin{equation}
H =\frac{1}{16\pi G} \int_{\Sigma_x}
[-s N {\cal{H}} + N^{\mu} {\cal{H}}_\mu ]
- \frac{1}{8 \pi G} \int_{S_x^\infty} [  N K^{(D-2)}
 - N^\nu p_{\mu \nu}r_{\nu} ] \ .
\end{equation}

Now, in order to define the physical Hamiltonian we must choose a
reference background.
Thus, given the space-time manifold $(M,g_{\mu \nu})$ we choose a
space-time manifold $(M,(g_0)_{\mu \nu})$ as the reference background
such that $g_{\mu\nu} = (g_0)_{\mu \nu}$ on the boundary $\Sigma^\infty$.%
\footnote{If we include matter fields $\phi_i$ we also need
that $(\phi_0)_i = \phi_i$ on $\Sigma^\infty$.}
The requirement that $g_{\mu\nu}$ and $(g_0)_{\mu\nu}$ coincide
on $\Sigma^\infty$ can be relaxed to an approximate agreement, provided
the approximation becomes exact as $\Sigma_\infty$ recedes to infinity.
We require furthermore that $X^\mu$ is a Killing vector
on $(M,(g_0)_{\mu \nu})$, i.e. that $\mathcal{L}_X (g_0)_{\mu \nu} = 0$.

Since $(M,(g_0)_{\mu \nu})$ is invariant under translations in $x$,
the momenta and the constraints vanish, and the physical Hamiltonian
$H - H_0$ becomes
\begin{equation}
H - H_0 =\frac{1}{16\pi G} \int_{\Sigma_x}
[-s N {\cal{H}} + N^{\mu} {\cal{H}}_\mu ]
- \frac{1}{8 \pi G} \int_{S_x^\infty}
\left[  N \left( K^{(D-2)} - K_0^{(D-2)} \right)
 - N^\nu p_{\mu \nu}r_{\nu} \right] \ ,
\end{equation}
where $K_0^{(D-2)}$ is the extrinsic curvature of $S^\infty_x$ in
the reference space $(M,(g_0)_{\mu \nu})$, with the appropriate choice
of labelling of slices so that $N_0 = N$.

We then define the energy or tension associated to translations in $x$
(time or space) as
\begin{equation}
\label{enten}
- \frac{1}{8 \pi G} \int_{S_x^\infty} [  N (K^{(D-2)}-K_0^{(D-2)})
 - N^\nu p_{\mu \nu} r_{\nu} ] \ .
\end{equation}

%%%%%%%%%%%%%%%%%%%%%%%%%%%%%%%%%%%%%%%%%%%%%%%%%%%%%%%%%%%%%
\subsection{General formulas for energy and tension}
\label{genform}

We now write down the general formulas for
energy and tension that follow from \eqref{enten} in more
detail.

Consider first a time-like $X^\mu$, which means
that $s=-1$, i.e. the submanifolds $\{ \Sigma_x \}$
are Euclidean. Then $x$ is a time,
and we can then define the energy/mass for
the space-time $(M,g_{\mu \nu})$ with respect
to the static space-time $(M,(g_0)_{\mu \nu})$ as
\begin{equation}
\label{energ}
E = - \frac{1}{8 \pi G} \int_{S_x^\infty} [  N (K^{(D-2)}-K_0^{(D-2)})
 - N^\nu p_{\mu \nu} r_{\nu} ] \ ,
\end{equation}
where $X^\mu = N n^\mu + N^\mu$.
Note that $(M,g_{\mu \nu})$ is required to be approximately
static at infinity in this case.
This formula \eqref{energ} is of course the Hawking-Horowitz
formula of \cite{Hawking:1996fd}.

Consider instead now a space-like $X^\mu$, which means that
$s=1$, i.e. that the submanifolds $\{ \Sigma_x \}$ are Lorentzian.
Then $x$ is a space direction and we can define the tension using
\eqref{enten}.
However, we would like to cancel out the integration over the time.
Assuming that the space-time $(M,g_{\mu \nu})$ is stationary
we can consider some time-function $t \in [t_{\rm min},t_{\rm max}]$
for $M$ defining a time-interval $\Delta t = t_{\rm max}-t_{\rm min}$.
Thus, dividing by the time interval $\Delta t$ will cancel the
integration over the time-function $t$, and we define
therefore the tension as
\begin{equation}
\label{tens}
\mathcal{T}
= - \frac{1}{\Delta t} \frac{1}{8 \pi G} \int_{S_x^\infty} [  F (K^{(D-2)}-K_0^{(D-2)})
 - F^\nu p_{\mu \nu} r_{\nu} ] \ ,
\end{equation}
where we have used the notation $X^\mu = F n^\mu + F^\mu$.
Note that $(M,g_{\mu \nu})$ is required to be approximately
translationally invariant along $x$ at infinity in this case.
The tension $\mathcal{T}$ in \eqref{tens} is the tension along
the $x$ direction.

%%%%%%%%%%%%%%%%%%%%%%%%%%%%%%%%%%%%%%%%%%%%%%%%%%%%%%%%%%%%%%
\section{Asymptotically-flat space-times \label{secflat} }

In this section we first explain that the general results
\eqref{energ}, \eqref{tens} reduce to the known ADM-type formulas for
mass and tension in asymptotically-flat space-times.
We then present in Section \ref{secfirstlaw} the general Smarr
formula and the first law of thermodynamics.

%%%%%%%%%%%%%%%%%%%%%%%%%%%%%%%%%%%%%%%
\subsection{Total mass and tension for asymptotically-flat space-times}

In Ref.~\cite{Hawking:1996fd} it was shown that the expression
\eqref{energ} for the mass correctly reduces to the ADM mass
for asymptotically-flat space-times \cite{Arnowitt:1962}
\begin{equation}
\label{admmass}
M  = \frac{1}{16 \pi G} \int_{S_t^\infty} d S^m
(\partial^n h_{mn} - \delta^{nl} \partial_m h_{nl}) \ .
\end{equation}
Here $g^{(t)}_{mn} = \delta_{mn} + h_{mn}$
is the $(D-1)$-dimensional metric as defined in
\eqref{defgx} when foliating the space-time along the time $t$.
We assume for simplicity here that the background metric is $\delta_{mn}$.
The indices $m,n,l$ run over the $D-1$ directions perpendicular to $t$.

In Appendix \ref{redux} we show that 
for an asymptotically-flat space-time 
the general tension formula \eqref{tens}
reduces to the ADM-type formula
\begin{equation}
\label{admtens}
\CT = \frac{1}{16 \pi G}  \int_{\hat{S}_z^\infty} d S^m
(\partial^n h_{mn} - \eta^{nl} \partial_m h_{nl}) \ .
\end{equation}
Here $g^{(z)}_{mn} = \eta_{mn} + h_{mn}$ is the $(D-1)$-dimensional metric
as defined in
\eqref{defgx} when foliating the space-time along the spatial direction $z$.
We assume for simplicity here that the background metric is
$\eta_{mn} = \mbox{diag} (-1,1,...,1)$.
The indices $m,n,l$ run over the $D-1$ directions perpendicular to $z$.
Note that in going from \eqref{tens} to
\eqref{admtens} we have cancelled the $\Delta t$ factor by integrating
over the time at infinity $\int dt = \Delta t$, so that the
remaining integration is over $\hat S_z^\infty = S_z^\infty/ \Delta t$.
The tension formula \eqref{admtens} agrees with the one
in \cite{Traschen:2001pb,Townsend:2001rg}.

The equations above provide a nice set of covariant expressions for
the mass and tension(s) in asymptotically-flat space-times with
one (or more) space-like isometries at infinity. However, by
solving Einstein equations at infinity where the gravitational field
is weak and using the principle of equivalent sources%
\footnote{The principle of equivalent sources is the principle that any
source of gravitation affecting the asymptotic region the same way
should also have the same values for the physical parameters
associated with the sources of the gravitational field.}
it is possible
to reduce to even simpler expressions. The derivation, which builds
on \cite{Townsend:2001rg,Harmark:2003dg,Kol:2003if} is given  in Appendix
\ref{apptension}.

The result is derived for the case of a $D$-dimensional
space-time with coordinates
$t$, $z^a$, $a=1 ,\ldots, k$, $x^i$, $i = 1 ,\ldots, D-k-1$ and
$r^2 = \sum_{i=1}^{D-k-1} (x^i)^2$ is the radial
coordinate in the transverse space. Beyond
the time translation Killing vector $\partial/\partial t$,
we assume in addition that  $\partial/\partial z_a$ are
asymptotic Killing vectors. We consider each of the directions
$z^a$ to be compactified on a circle with circumference $L_a$.
Finally, the space is taken to be asymptotically flat, so that the
spatial part of the metric at infinity is
 $\R^{D-k-1} \times \T^k  $, 
where $\T^k \equiv (S^1)^k$ is a rectangular $k$-torus
with volume $V_k = \prod_{a=1}^k L_a$.

The mass $M$ and tensions $\CT_a$ in each of the compact directions
 are then given by%
\footnote{Here $\Omega_l = 2\pi^{\frac{l+1}{2}}/\Gamma \left( 
\frac{l+1}{2} \right)$ is the volume of the unit $l$-sphere.}
\begin{equation}
\label{findM}
M = \frac{   V_k \Omega_{D-k-2}}{16 \pi G}
\left[ (D-k-2) c_t - \sum_a c_a \right] \ ,
\end{equation}
\begin{equation}
\label{findT}
\CT_a = \frac{  V_k \Omega_{D-k-2}}{16 \pi G L_a}
\left[ c_t - (D-k-2) c_a -\sum_{b\neq a} c_a\right] \spa a = 1,\ldots, k \ ,
\end{equation}
where $c_t$ and $c_a$ are the leading corrections of the metric at infinity
\begin{equation}
\label{leadmetric}
g_{tt} = -1 + \frac{c_t}{r^{D-k-3}} \spa g_{aa}
= 1 + \frac{c_a}{r^{D-k-3}} \spa
a = 1, \ldots, k \ .
\end{equation}
These equations enable one to measure
the mass \eqref{admmass} and tension \eqref{admtens}
for the class of space-times defined above
in terms of the asymptotics of the metric.

It is important to realize that not all values of $M$ and $\CT_a$
correspond to physically reasonable matter. Just like we demand
that $M \geq 0$, we also have bounds on the tensions $\CT_a$.
In the general framework of this section we have the bounds
\begin{equation}
\label{bounds}
M \geq \frac{1}{D-3} \sum_a L_a \CT_a \spa \CT_a \geq 0  \ .
\end{equation}
The first bound follows from the Strong Energy Condition
which basically states that any physically sensible
source for the gravitational field can not be repulsive.
In terms of the energy-momentum tensor
it is the statement that $T_{00} + \frac{1}{D-2} T^\mu_{\ \mu} \geq 0$.
Integrating this relation over the space-like directions and
using \eqref{zeroT}, \eqref{bulk} then gives the stated result.

The bound $\CT_a \geq 0$ (for a given $a$) corresponds
instead to the statement that the tension can not be
negative, as proven in \cite{Traschen:2003jm,Shiromizu:2003gc}.

%%%%%%%%%%%%%%%%%%%%%%%%%%%%%%%%%%%%%%%%%%%%%%%%%%%%%%%%%%%
\subsection{Smarr formula and first law of thermodynamics}
\label{secfirstlaw}

We can push our general considerations further by splitting up 
the energy-momentum tensor as the sum
\begin{equation}
\label{Tsum}
T_{\mu \nu} = T^{\rm gr}_{\mu \nu} + T^{\rm mat}_{\mu \nu} \ ,
\end{equation}
where $T^{\rm gr}_{\mu \nu}$ is the energy-momentum tensor of the 
gravitational field
relative to  the $D$-dimensional flat background and
$T^{\rm mat}_{\mu \nu}$ the contribution of additional matter, such as
scalar fields or $p$-form gauge potentials (see also Appendix
\ref{apptension}). 

Using now the principle of equivalent sources, we assume in the
following that our space-time is everywhere nearly flat.
We can then write the mass $M$ and tension $\CT_a$ as
\begin{equation}
\label{bulk}
M = \int d^{D-1}x \, T_{00} \spa  \CT_a = -\frac{1}{L_a}\int d^{D-1} x\, T_{aa} \spa
a = 1 ,\ldots, k \ .
\end{equation}
where $L_a$ is the size of the $a$'th direction.

Using instead the matter energy-momentum tensor $T^{\rm mat}_{\mu \nu}$
we can write the matter contributions
to the mass and tension denoted by $M^{\rm mat}$ and $\CT_a^{\rm mat}$
respectively, as
\begin{equation}
\label{bulkmat}
M^{\rm mat} = \int d^{D-1}x \, T^{\rm mat}_{00} \spa
\CT_a^{\rm mat} = -\frac{1}{L_a}\int d^{D-1}x \, T_{aa}^{\rm mat} \spa
a = 1 ,\ldots, k \ .
\end{equation}

Consider now again the class of space-times for which \eqref{findM},
\eqref{findT} was derived, and assume in addition the presence of
 a single connected event horizon.
Then, following \cite{Townsend:2001rg,Harmark:2003dg,Kol:2003if}, it is
possible to derive a generalized Smarr formula. For completeness, the
derivation is given in Appendix \ref{Smarrproof}.
Here we present the general Smarr formula
 \eqref{Smarrgen} in the simple form
\begin{equation}
\label{Smarrgen2}
(D-2) T S  =
\left( D-3 - \sum_a n_a \right) (M-M^{\rm mat}) \ ,
\end{equation}
where we have introduced the {\sl relative tensions}
\begin{equation}
\label{ndef}
n_a \equiv
\frac{L_a(\CT_a -\CT_a^{\rm mat})}{M - M^{\rm mat}}  \spa a= 1,\ldots, k \ ,
\end{equation}
generalizing the relative tensions introduced in \cite{Harmark:2003dg}.%
\footnote{Note that the relative tension $n_a$ along a direction
was denoted {\sl relative binding energy}
in \cite{Harmark:2003dg} since the tension also can be
viewed as a binding energy along the particular direction. Here we
adopt a different nomenclature since it is conventional to denote
the quantity measured in \eqref{tens} as the tension.}
Note that \eqref{Smarrgen2}, \eqref{ndef} depend only on the gravitational
contribution
to the mass and tensions.

We also give the generalized first law of thermodynamics
for this class of space-times,
\begin{equation}
\label{firstlaw}
\delta M = T \delta S + \sum_a \CT_a^{\rm eff} \delta L_a \ ,
\end{equation}
showing that the effective tension in each compact direction
\begin{equation}
\label{efftens}
\CT_a^{\rm eff} = \CT_a - \CT_a^{\rm mat} \ ,
\end{equation}
is governed by the gravitational part of the tension.
The first law can be for example obtained by combining
 the Smarr formula \eqref{Smarrgen2}, with scaling relations and
 using  Euler's theorem \cite{Townsend:2001rg}. It also correctly reduces
 to the first law obtained in \cite{Harmark:2003dg,Kol:2003if} for the
 case of neutral black objects on the cylinder.

Finally, we note that while we still impose the bounds \eqref{bounds}
on the tensions $\CT_a$, we can in fact now use the above to obtain
stronger bounds on our physical variables.
This is because the
energy-momentum tensor in \eqref{Tsum} is a sum of a gravitational part
and a matter
part and we require these bounds to hold independently
for both of the two contributions.
The reason for this is that we expect both the matter part
and the gravitational part to behave in a physically sensible
fashion.
Since the matter usually obeys
the bounds by construction we can extract some stronger bounds
on our physical variables by considering the gravitational part
by itself.
The bounds on the gravitational part are
$(D-3) (M - M^{\rm mat}) \leq \sum_a L_a (\CT_a - \CT_a^{\rm mat})$,
and $\CT_a - \CT_a^{\rm mat} \geq  0  $. Using the definition \eqref{ndef}
we can write these as
\begin{equation}
\label{nbound}
\sum_a n_a \leq D-3 \spa n_a \geq 0 \ .
\end{equation}
As stated above, these bounds are stronger than the bounds
\eqref{bounds}.

%%%%%%%%%%%%%%%%%%%%%%%%%%%%%%%%%%%%%%%%%%%%%%%%%%%%%%%%%%%%%%
\section{Examples \label{secex} }

In this section we present two simple examples that illustrate the general
definition of gravitational tension.

\subsection{Kaluza-Klein black holes}

As the first example we consider vacuum solutions of higher-dimensional
General Relativity (i.e. pure gravity) for the Kaluza-Klein type space-times
 $\CM^d \times S^1$ with $d \geq 4$
where $\CM^d$ is the $d$-dimensional Minkowski space-time.
More precisely, we consider all static vacuum solutions that asymptote
to $\CM^d \times S^1$ and that have an event horizon.
We call these solutions Kaluza-Klein black holes here.

Define the Cartesian coordinates for $\CM^d$ as
$t,x^1,...,x^{d-1}$ along with the radius $r = \sqrt{(x^1)^2 + \cdots
+ (x^{d-1})^2}$. Let moreover $z$ be the coordinate of the $S^1$
with period $L$. Since the Kaluza-Klein black holes are asymptotically
flat, we may use the expressions \eqref{findM}-\eqref{leadmetric}
to find the mass $M$ and tension $\CT$ along the compact $z$-direction.
It then easily follows from these equations that for
any given Kaluza-Klein black hole solution we can write
\cite{Harmark:2003dg,Kol:2003if}
\begin{equation}
M = \frac{\Omega_{d-2} L}{16 \pi G}
\left[ (d-2) c_t - c_z \right]
\spa
\CT = \frac{\Omega_{d-2}}{16 \pi G}
\left[ c_t - (d-2) c_z \right] \ ,
\end{equation}
where $c_t$, $c_z$ are read off from the metric components
\begin{equation}
g_{tt} = -1 + \frac{c_t}{r^{d-3}}
\spa
g_{zz} = 1 + \frac{c_z}{r^{d-3}} \ ,
\end{equation}
for $r \rightarrow \infty$.

We also immediately recover from \eqref{Smarrgen2}, \eqref{firstlaw}
the Smarr relation and first law of thermodynamics in  this case
\cite{Harmark:2003dg,Kol:2003if}
\begin{equation}
(d-1) TS = (d-2-n) M \spa
\delta M = T \delta S + \CT \delta L \ ,
\end{equation}
where $n=L \CT/M$ is the relative tension. Finally, the
general relation \eqref{nbound}
implies the bounds \cite{Harmark:2003dg}
\begin{equation}
 0 \leq n \leq d-2 \ ,
 \end{equation}
on the relative tension.

%%%%%%%%%%%%%%%%%%%%%%%%%%%%%%%%%%%%%%%%%%%%%%%%%%
\subsection{Near-extremal $p$-branes}

The second example we consider is that of near-extremal $p$-branes
of type IIA/B String theory and M-theory. As these branes are not
asymptotically flat, this provides a nice application of
the general expression \eqref{tens} for gravitational tension.

The near extremal $p$-brane solution
in $D=1+p+d$ dimensions  has the form
\begin{equation}
\label{nearsol1}
ds ^2
= \hat{H}^{-\frac{d-2}{D-2}} \left( - f dt^2
+ \sum_{k=1}^p (du^k)^2 + \hat{H} \left[ f^{-1} dr^2
+   r^2 d\Omega_{d-1}^2 \right] \right) \ ,
\end{equation}
\begin{equation}
\label{nearsol3}
 e^{2\phi} = \hat{H}^a
\spa
A_{(p+1)} = \hat{H}^{-1} dt \wedge du^1 \wedge \cdots \wedge du^p \ ,
\spa
f = 1 - \frac{r_0^{d-2}}{r^{d-2}} \spa
\hat{H} = \frac{\hat h_d}{r^{d-2}} \ ,
\end{equation}
where the metric is written in
the Einstein frame. 
Here $\phi$ is the dilaton (for $D=10$),
$A_{(p+1)}$ the gauge potential and $a$ is a number determining
the coupling of the $A_{(p+1)}$ gauge potential to the dilaton.
The background asymptotes to the
near-horizon limit of 
the 1/2 BPS extremal $p$-brane of String/M-theory 
corresponding to \eqref{nearsol1}-\eqref{nearsol3} with $r_0= 0$, 
which is taken as the reference background when computing the energy and
tension.

The relevant tensions in this case are those corresponding to the
$p$ spatial world-volume directions which we take to be compactified on
equal size circles of circumference $L$, so that the total volume
$V= L^p$. 
Therefore the tension $\CT_k = \CT $ 
is the same in each direction $u^k$, $k= 1 ,\ldots, p$.

Although the Hawking-Horowitz energy computation is well known for these
branes we present for clarity the computation of energy $E$
 and tension  $\CT$ in
parallel below. Note that as we are using the extremal $p$-brane
as the reference background, the expressions \eqref{energ}, \eqref{tens}
 are of course the energy and tension above extremality.
To compute these quantities  we first recall that
the extrinsic curvature $K^{(D-2)}$ can for our purposes
be written as
\begin{equation}
\label{excur}
K^{(D-2)} = \frac{1}{\sqrt{g_{rr}}} \partial_r \log \sqrt{|g^{D-2}|} \ .
\end{equation}
Here $g^{D-2}$ is the determinant of the metric obtained
from \eqref{nearsol1} by omitting $t,r$ when computing $E$ and omitting
$u^k,r$ (for a particular $k$)
when computing $\CT$. We denote the corresponding
expressions by $K_t$ and $K_u$ respectively.

It is then not difficult to use \eqref{excur} for the metric
\eqref{nearsol1} and compute
\begin{equation}
\label{excurt}
K_t = \sqrt{f(r_m)} \hat{H}(r_m)^{-\frac{p+1}{2(D-2)}} 
\frac{d}{2 r_m}     \ ,
\end{equation}
\begin{equation}
\label{excurz}
K_u = \sqrt{f(r_m)} \hat{H}(r_m)^{-\frac{p+1}{2(D-2)}} 
\left[ \frac{d}{2} + \frac{d-2}{2}
\left( \frac{r_0}{r_m} \right)^{d-2} \right]\frac{1}{r_m} \ ,
\end{equation}
where we used $r_m \gg r_0$ since $r_m$ is sent to infinity
in the end. The corresponding expressions for
the extrinsic curvature of the reference space in both cases are
obtained from \eqref{excurt}, \eqref{excurz} by setting $r_0=0$ 
and substituting $r_m$ with $r_{\rm eff}$.
Here $r_{\rm eff}$ is the radius corresponding to $r_m$
for the reference space, defined so that the
metrics for the $(D-2)$-dimensional subspaces are equal. 
The relation between $r_m$ and $r_{\rm eff}$ is then
obtained by imposing that the radius of the $S^{d-1}$ in the brane
space-time is equal to that of the radius of the $S^{d-1}$ in
the reference space. In the present case, this simply means that
$r_{\rm eff} = r_m$ to leading order.
Finally, we need for each case the lapse functions and integration
measures: $N = (g_{tt}^{(0)})^{1/2} = \hat H^{-\frac{d-2}{2(D-2)}}$,
$\sqrt{h} = \sqrt{\hat H} r_m^{d-1}$ for the energy
and $F = (g_{zz}^{(0)})^{1/2} = \hat H^{-\frac{d-2}{2(D-2)}}$,
$\sqrt{h} = \sqrt{\hat H}  r_m^{d-1}$ for the tension.

Using all this in \eqref{energ}, \eqref{tens}
we have the final results
\begin{equation}
E = \frac{d}{2} \frac{V  \Omega_{d-1}}{16 \pi G} r_0^{d-2} \spa
L \CT  = - \frac{d-4}{2}\frac{V  \Omega_{d-1}}{16 \pi G} r_0^{d-2} \ ,
\end{equation}
for the energy and spatial world-volume tension of a near-extremal
$p$-brane.
Note that the quantity $-E + p \, L\CT$ is the trace of the (integrated)
energy-momentum tensor on the brane, which is zero only for
the non-dilatonic (D3, M2, M5) branes.
We remark that the tension was calculated via a different method in
Refs.~\cite{Myers:1999ps,Cai:1999xg} (see also e.g.
Refs.~\cite{Awad:1999xx,DeBenedictis:1999yn}).

It is actually more natural to use instead of tension
the pressure given by
\begin{equation}
\label{press}
P= - \frac{L \CT}{V} =\frac{d-4}{2}\frac{ \Omega_{d-1}}{16 \pi G} r_0^{d-2} \ .
\end{equation}
The pressure is positive for $d > 4$, which includes the near-extremal
D3, M2 and M5 brane. In these cases, we can interpret the pressure
directly in terms of the massless excitations of the corresponding field
theory at finite temperature.

To consider the thermodynamics of the near-extremal $p$-brane we use
the well-known results for temperature and entropy
\begin{equation}
T = \frac{d-2}{4\pi} \hat h_d^{-\frac{d-2}{2}} r_0^{\frac{d-4}{2} }
\spa S = \frac{V \Omega_{d-1}}{4 G} \hat h_d^{\frac{d-2}{2}}
r_0^{\frac{d}{2}} \ .
\end{equation}
It is then easy to show that the tension is directly related to
the Helmholtz free energy
\begin{equation}
L \CT = F  \spa F = E - TS \ .
\end{equation}
Together with the identification of the pressure \eqref{press} this
implies that the Gibbs free energy vanishes for near-extremal
$p$-branes
\begin{equation}
G = 0 \spa G = E - TS + P V \ .
\end{equation}
It is actually not difficult to verify that any conformally invariant
theory in $p+1$ dimensions has $G=0$. Here, however, we see that it
holds for all near-extremal $p$-brane theories, regardless whether they
are conformal or not. Finally, we note that it is a simple task to
explicitly check that the first law $\delta E = T \delta S -P \delta V$
is satisfied using the pressure in \eqref{press}.

%%%%%%%%%%%%%%%%%%%%%%%%%%%%%%%%%%%%%%%%%%%%%%%%%%%%%%%%%%%%%%
\section*{Acknowledgments}

We thank Simon Ross for useful discussions.

%%%%%%%%%%%%%%%%%%%%%%%%%%%%%%%%%%%%%%%%%%%%%%%%%%%%%%%%%%%%%%%%%%%
\begin{appendix}

%%%%%%%%%%%%%%%%%%%%%%%%%%%%%%%%%%%%%%%%%%%%%%%%%%%%%%%%%%%%%%%
\section{Reduction to ADM-type formula}
\label{redux}

We show in this appendix how the general tension formula \eqref{tens}
reduces to the ADM-type formula \eqref{admtens} for asymptotically-flat
space-times. 
Write $g^{(z)}_{mn}= \eta_{mn} + h_{mn}$ as the metric
on the $(D-1)$-dimensional subspace transverse to the $z$-direction.
Since the reference space-time is flat we can choose the
metric of the reference space-time as $\eta_{mn}$. 
This means that $h_{mn}$ should be zero on $S^\infty_z$
for the tangential directions.

We now first consider the case of $S^\infty_z$ being a hyper-cube.
In the flat reference background each side of the hyper-cube
is defined by one of the $2D-2$ relations 
$x^i = \pm \alpha$, $i=1,...,D-1$. 
$\alpha$ is then sent to infinity in the end.
Then \eqref{tens} becomes
\begin{equation}
\label{interm}
\CT = - \frac{1}{\Delta t} \frac{1}{8\pi G} \int_{S^\infty_z}
d^{D-2} x \left( K^{(D-2)} - K_0^{(D-2)} \right) \ .
\end{equation}
We focus below without loss of generality on the side $x^1=\alpha$.
For the reference space-time the unit normal vector is simply
$(r_0)^m = \delta_{m1}$.
For the asymptotically-flat space-time it is instead
$r^m = \delta_{m1} - \frac{1}{2} \eta^{mn} h_{n1}$.%
\footnote{This is because $x^1 = \alpha$ is not the equation
for the hypersurface
in the asymptotically-flat space-time since the metric is not
the Minkowski metric.
To find the normal vector notice that we have the inverse Vielbeins
$e^m_a = \delta^m_a - \frac{1}{2} \eta^{mn} h_{na}$
so that $g^{mn} = \eta^{ab} e^m_a e^n_b$. 
From the fact that the Vielbeins precisely relate a Minkowski metric
to the asymptotically flat metric,
we get then that $r^m = e^m_a (r_0)^a = e^m_1 = \delta_{m1} 
- \frac{1}{2} \eta^{mn} h_{n1}$.}
We have then
\begin{equation}
K^{(D-2)} = D^{(z)}_m r^m = \partial_m r^m + \Gamma^m_{mn} r^n
= \frac{1}{2}  \left[ -  \partial^n h_{nm} 
+ \eta^{nl} (\partial_m h_{nl} ) \right] r^m \ ,
\end{equation}
where we used that $\partial_m r^m = - \frac{1}{2} \partial^n h_{n1}
= - \frac{1}{2} \partial^n h_{nm} r^m$ and that 
$\Gamma^m_{mn} = \frac{1}{2} \eta^{ml} (\partial_n h_{ml} )$
to first order in $h_{mn}$.
Since $K_0^{(D-2)} = \partial_m (r_0)^m = 0$ we get therefore
\begin{equation}
\label{extreq}
K^{(D-2)} - K_0^{(D-2)} = \frac{1}{2}  \left[ -  \partial^n h_{nm} 
+ \eta^{nl} (\partial_m h_{nl} ) \right] r^m \ .
\end{equation}
Eq.~\eqref{extreq} clearly holds for all the sides of the hyper-cube.
Therefore, \eqref{interm} gives
\begin{equation}
\CT = \frac{1}{\Delta t} \frac{1}{16\pi G} \int_{S^\infty_z}
d^{D-2} x \left( \partial^n h_{nm}
- \eta^{nl} \partial_m h_{nl} \right) r^m \ ,
\end{equation}
which clearly is the same as \eqref{admtens}.
One can now generalize to other hypersurfaces $S^\infty_z$.
First we can generalize to surfaces build up by rectangular surfaces
of the type $x^m = \mbox{constant}$. For a general hypersurface $S^\infty_z$
we can then find surfaces built out of rectangular surfaces
that are arbitrarily close to $S^\infty_z$, thus proving that
it works for general surfaces also.

%%%%%%%%%%%%%%%%%%%%%%%%%%%%%%%%%%%%%%%%%%%%%%%
\section{Mass and tension from linearized Einstein equation}
\label{apptension}

In this appendix we solve the linearized Einstein equations for
 a static distribution of matter localized on the 
$\R^{D-k-1} $ times a $k$-torus
$ \T^k$. The derivation builds on that of 
Refs.~\cite{Townsend:2001rg,Harmark:2003dg,Kol:2003if}.

We first define for any energy-momentum tensor $T_{\mu \nu}$ the
tensor
\begin{equation}
\label{Sdef1}
S_{\mu \nu} \equiv  T_{\mu \nu} -
\frac{1}{D-2} \eta_{\mu \nu} T \spa T= T_{\mu}{}^{\mu}\ .
\end{equation}
Then, by eliminating the Ricci scalar $R$, the Einstein equations 
with the matter energy-momentum tensor 
$T_{\mu \nu}^{\rm mat}$ can be written in the form
\begin{equation}
R_{\mu \nu} = 8 \pi G  S_{\mu \nu}^{\rm  mat} \ ,
\end{equation}
where $S_{\mu \nu}^{\rm  mat}$ is related to $T_{\mu \nu}^{\rm mat}$
via \eqref{Sdef1}.
In the linearized approximation $g_{\mu \nu} = \eta_{\mu \nu}
+ h_{\mu \nu}$ we have that
\begin{equation}
R_{\mu \nu} = R^{(1)}_{\mu \nu} - 8 \pi G S_{\mu \nu}^{\rm gr} \ ,
\end{equation}
where
\begin{equation}
\label{linEin0}
 R^{(1)}_{\mu \nu} = -\frac{1}{2}[\square h_{\mu \nu} + (h_\lambda{}^\lambda)_{,\mu \nu}
-  (h_\mu{}^\lambda)_{,\nu \lambda} -  (h_\nu{}^\lambda)_{,\mu \lambda} ] \ ,
\end{equation}
and $S_{\mu \nu}^{\rm gr}$ determined via \eqref{Sdef1} by the energy-momentum tensor contribution $T_{\mu \nu}^{\rm gr}$ of the gravitational field relative to the $D$-dimensional Minkowski background. 
Note that $S_{\mu \nu}^{\rm gr}$ collects all higher-order terms 
of $R_{\mu \nu}$.
In the weak-field approximation the Einstein equations
thus reduce to the system
\begin{equation}
\label{linEin}
\square h_{\mu \nu} + (h_\lambda{}^\lambda)_{,\mu \nu}
-  (h_\mu{}^\lambda)_{,\nu \lambda} -  (h_\nu{}^\lambda)_{,\mu \lambda}
= - 16 \pi G S_{\mu \nu} \spa S_{\mu \nu} = S_{\mu \nu}^{\rm gr}
+ S_{\mu \nu}^{\rm  mat} \ ,
\end{equation}
and $S_{\mu \nu}$ depends on the total energy-momentum
tensor,  given by the sum of the gravitational and matter part,
$T_{\mu \nu} = T_{\mu \nu}^{\rm gr} + T_{\mu \nu}^{\rm mat}$.

Consider now a $D$-dimensional space-time with $k$ compact directions
$z^a$, $a=1 ,\ldots, k$ of periods $L_a$, and $D-k-1$ transverse
directions $x^i$, $i=1,\ldots, D-k-1$. The
spatial part is thus of the from $\R^{D-k-1} \times \T^k$, where $\T^k
= (S^1)^k$ is a rectangular $k$-torus. The flat metric thus reads
\begin{equation}
\label{flat}
ds^2 = - dt^2 + \sum_{a=1}^k (dz^a)^2 + \sum_{i=1}^{D-k-1} (dx^i)^2
= - dt^2 + \sum_{a=1}^k (dz^a)^2 + dr^2 + r^2 d \Omega_{D-k-2}^2 \ ,
\end{equation}
where we have defined $r^2 = \sum_{i=1}^{D-k-1} (x^i)^2$.

We consider a static distribution of matter localized on
 $\R^{D-k-1}$, and assume
 a diagonal energy-momentum tensor $T_{\mu \nu}$ with non-zero components
\begin{equation}
\label{Tnonzero}
T_{00}   \spa T_{aa}  \spa T_{ii} \ .
\end{equation}
Here $T_{00}$ is a function of $(z^a,x^i)$, while a given
$T_{aa}$ depends on all these coordinates except $z^a$, because of
energy-momentum conservation.
We also assume that the resulting space-time metric is asymptotic to the flat
metric \eqref{flat}, so that $T_{\mu \nu}$ vanishes in the 
limit $r \rightarrow \infty$ 
(sufficiently rapid). Now, energy-momentum conservation implies
$\sum_i \partial_i T_{ij} =0$ so that we can use partial integration
to write
\begin{equation}
\label{zeroT}
  \int d^{D-1 }x  T_{ii} =   \int_{S_\infty   } d S_j
x^i  T_{ij} = 0 \ ,
\end{equation}
where the first step uses $T_{ii} = \partial_j (x^i T_{ij})$ along
with Gauss law ($S_\infty = S^{D-k-2} \times \T^k $) and
in the last step we used the vanishing of $T_{ij}$ at spatial
infinity in the transverse space. As a consequence of \eqref{zeroT}
we find in particular that the bulk integrals
$\int d^{D-1}x T_{ii} =0$ for the non-zero components $T_{ii}$
in \eqref{Tnonzero}.

To obtain the relation between the total energy $M$ and tensions $\CT_a$ in
the compact directions defined by \eqref{bulk} and
the asymptotics of the metric,  we now solve the linearized Einstein
equations \eqref{linEin} for the energy-momentum tensor \eqref{Tnonzero}.
We use the fact that the metric perturbation $h_{\mu \nu}$ is time-independent,
along with the fact that it is asymptotically diagonal.
Then  \eqref{linEin} reduce to the set of equations
\begin{equation}
\label{eq0a}
\nabla^2 h_{00} = -  16 \pi G S_{00} \spa
\nabla^2 h_{aa} = -  16 \pi G S_{aa} \spa a = 1 ,\ldots, k \ ,
\end{equation}
\begin{equation}
\label{eqi}
-2 \sum_{i,j=1}^{D-k-1} (h_{ij,ij} -h_{ii,jj} )
-\nabla^2 h_{00} + \sum_a \nabla^2 h_{aa} = - 16 \pi G \sum_{i=1}^{D-k-1} S_{ii} \ ,
\end{equation}
where $\nabla^2$ is the Laplacian operator on $\R^{D-k-1} \times \T^k$.

We focus first on the $1 + k$ equations in \eqref{eq0a}. Using
the definitions for the mass $M$ and binding energies $\CT_a$  in
 \eqref{bulk}, \eqref{zeroT} as well as \eqref{Sdef1} and \eqref{zeroT},
 these are easily solved to leading order yielding
\begin{equation}
\label{M1}
M = -\frac{1}{D-k-3} \frac{1}{16 \pi G} \int_{S_\infty} d S_i
[ (D-k-2)h_{00,i} - \sum_a h_{aa,i} ] \ ,
\end{equation}
\begin{equation}
\label{T1}
L_a \CT_a = -\frac{1}{D-k-3} \frac{1}{16 \pi G} \int_{S_\infty} dS_i
[h_{00,i}  - (D-k-2) h_{aa,i} - \sum_{b \neq a} h_{bb,i} ] \ .
\end{equation}
These are surface integrals over $S_\infty$, which is the
$(D-k-2)$-sphere at infinity times the $k$-torus $\T^k$.
 From these we obtain the expressions \eqref{findM}, \eqref{findT} in
 terms of the leading corrections \eqref{leadmetric} of the metric at
 infinity.

It remains to show that $M$ and $\CT_a$ in \eqref{M1}, \eqref{T1}
(and hence the definitions in \eqref{bulk}, \eqref{zeroT})
actually agree with
the ADM formulae \eqref{admmass}, \eqref{admtens} obtained for the asymptotically
flat case from the general Hamiltonian treatment. For the case at
hand, these can be written as
\begin{equation}
\label{MA}
M^{(\rm ADM)} = \frac{1}{16 \pi G} \int_{S_\infty} d S_i
[h_{ij,j} - h_{jj,i} - h_{aa,i} ] \ ,
\end{equation}
\begin{equation}
\label{TA}
L_a \CT_a^{(\rm ADM)} = \frac{1}{16 \pi G} \int_{S_\infty} dS_i
[h_{ij,j}  - h_{jj,i} +  h_{00,i} - \sum_{b \neq a} h_{bb,i} ] \ ,
\end{equation}
where we remind the reader that repeated indices are summed over.
To this end, we use the linearized Einstein equations \eqref{eqi} in the
transverse space. Using Eq.~\eqref{eq0a} we can rewrite this relation as
\begin{equation}
\label{cons}
\frac{1}{16\pi G} (h_{ij,ij} -h_{ii,jj}  -h_{aa,ii} )
= \frac{1}{2} ( S_{00} + S_{aa} + S_{ii} ) \ .
\end{equation}
Integrating this equation over all spatial dimensions, we see that
the left side is the ADM mass formula \eqref{MA}. On the other hand,
for the right side we may use \eqref{Sdef1}, \eqref{bulk} and \eqref{zeroT}
 to show that
$\frac{1}{2}\int d^{D-1}x \, (S_{00} + S_{aa} + S_{ii}) =  M$.
We thus conclude that the relation
\eqref{cons} tells us that $M= M_{\rm ADM}$.
Turning to the ADM tension \eqref{TA}, rearranging terms this can be written
as
\begin{eqnarray}
L_a \CT_a^{(\rm ADM)} & = &  \frac{1}{16 \pi G} \int dS_i [\partial_i (h_{00}  + h_{aa})
+  (h_{ij,j} - h_{jj,i}  - h_{bb,i} )] \nn \\
&=&  (-   M + L_a \CT_a )+ M_{\rm ADM}  = L_a \CT_a \ ,
 \end{eqnarray}
 where we used \eqref{M1}, \eqref{T1}, \eqref{MA} in the second step,
 and in the last step the fact that we already showed  $M= M_{\rm ADM}$. Thus the
 definitions of the tensions also agree, and we can omit the
 label ``ADM''.

%%%%%%%%%%%%%%%%%%%%%%%%%%%%%%%%%%%%%%%%%%%%%%%%%%%%%%%%%%%%%%%%
\section{Proof of generalized Smarr formula \label{Smarrproof} }

To derive the Smarr formula we consider the Komar integral
\begin{equation}
I_S = - \frac{1}{16 \pi G} \int_S dS_{\mu \nu} D^\mu \xi^\nu \ ,
\end{equation}
where $S$ is a $(D-2)$-dimensional hypersurface
 and $\xi$ is a Killing vector for the metric.
Consider now a static  solution on $\R^{D-k-1} \times \T^k$
with an event horizon.
Consider furthermore a certain time $t=t_0$.
Define $S_h$ to be the null-surface of the event horizon at $t=t_0$.
We also choose a
$(D-2)$ dimensional surface at $r=\infty$ for $t=t_0$,
which we call $S_\infty$, so that essentially $S_\infty = S^{D-k-2} \times \T^k$.
By Gauss theorem we have
\begin{equation}
I_{S_h} -I_{S_{\infty}}
= \frac{1}{16 \pi G_N} \left( \int_{S_{\infty}} dS_{\mu \nu} D^\mu \xi^\nu
- \int_{S_h} dS_{\mu \nu} D^\mu \xi^\nu \right)
= \frac{1}{8 \pi G_N} \int_{V} dS_\mu D_\nu D^\mu \xi^\nu \ ,
\end{equation}
where $V$ is the $(D-1)$-dimensional volume between $S_h$ and $S_\infty$
so that $\partial V = S_h \cup S_\infty$.
Now we use that for a Killing vector we have $D_\nu D^\mu \xi^\nu
= R^\mu_{\ \nu} \xi^\nu$ along with the fact that the sourced
Einstein equation  $R_{\mu \nu} = 8 \pi G S^{\rm mat}_{\mu \nu}$
holds everywhere in $V$, as we are away from the black object.
We thus have
\begin{equation}
\label{Komarrel}
I_{S_h} = I_{S_{\infty}} +\int_{V} dS_\mu (S^{\rm mat})^\mu {}_\nu
\xi^\nu \ .
\end{equation}

Since we have a static solution we can choose $\xi$ to be the
time-translation Killing vector field, i.e. $\xi = \partial / \partial t$.
We then compute \cite{Townsend:1997ku}
\begin{equation}
\label{ISH}
I_{S_h} = - \frac{1}{8 \pi G_N} \int_{S_h} dA k^\nu ( D_\nu k^\mu ) n_\mu
= - \frac{\kappa }{8 \pi G_N} \int_{S_h} dA k^\mu n_\mu
= \frac{\kappa A}{8 \pi G_N} = T S \ .
\end{equation}
On the other hand we have asymptotically
\begin{equation}
\label{ISI}
I_{S_{\infty}} =
- \frac{1}{16 \pi G_N} \int_{S_{\infty}} dS_{0r} \partial_r g_{00}
= \frac{1}{D-2} \left[ (D-3) M - \sum_a L_a \CT_a\right] \ .
\end{equation}
Here  we first used that
 the non-zero components of $D^\mu k^\nu$ are
$D^0 k^r = - D^r k^0 = \frac{1}{2} \partial_r g_{00}$ to leading order,
and \eqref{M1}, \eqref{T1} are used in the last step.
Finally we need to compute the last term in \eqref{Komarrel} which reads
\begin{eqnarray}
\int_{V} dS_0 (S^{\rm mat})^{0} {}_0  & = &  - \frac{1}{D-2}
\int d^{D-1}x \left[ (D-3) T_{00}^{\rm mat} + \sum_a L_a T_{aa}^{\rm mat}\right] \nn
\\
&=&- \frac{1}{D-2} \left[ (D-3) M^{\rm mat} - \sum_a L_a \CT_{a}^{\rm mat} \right] \ ,
\end{eqnarray}
where we recall the definition \eqref{Sdef1} of $S_{\mu \nu}$.
Putting it all together then yields the Smarr formula
\begin{equation}
\label{Smarrgen}
(D-3)(M - M^{\rm mat})  =  (D-2) T S +
\sum_a L_a (\CT_a- \CT_a^{\rm mat}) \ .
\end{equation}

\end{appendix}

%%%%%%%%%%%%%%%%%%%%%%%%%%%%%%%%%%%%%%%%%%%%%%%%%%%%%%%%%%%%%%%%%%%

\addcontentsline{toc}{section}{References}

%The following two lines is for bibtex only:
%\bibliographystyle{utphys}
%\bibliography{bibrot,biblioniels}
%\bibliographystyle{../INPUT/utphys}
%\bibliography{../BIB/bibrot,../BIB/biblioniels}

\providecommand{\href}[2]{#2}\begingroup\raggedright\endgroup

\end{document}